# Emergent quantum mechanics of the event-universe, quantization of events via Denrographic Hologram Theory

Oded Shor PhD,[1,3] * & Felix Benninger MD[1,2,3] & Andrei Khrennikov, PhD[4]

[1]Felsenstein Medical Research Centre, Petach Tikva, Israel; [2]Department of Neurology, Rabin Medical Centre, Petach Tikva, Israel; [3]Sackler Faculty of Medicine, Tel Aviv University, Tel Aviv, Israel; [4]Faculty of Technology, Department of Mathematics, Linnaeus University, Vaxjö, Sweden

Word count: 3900

Figures: 2

**Corresponding author (*):**

Oded Shor, Felsenstein Medical Research Centre, Petach Tikva, Israel; shor.oded@gmail.com

**Email of all authors:**

| | |
|---|---|
| Oded Shor | shor.oded@gmail.com |
| Felix Benninger | benninger@tauex.tau.ac.il |
| Andrei Khrennikov | Andrei.khrennikov@lnu.se |

**Running title:** Emergent quantum mechanics of the event-universe, quantization of events via Denrographic Hologram Theory

**Ethical Publication Statement**

We confirm that we have read the Journal's position on issues involved in ethical publication and affirm that this report is consistent with those guidelines.

**Funding**

This research did not receive any specific grant from funding agencies in the public, commercial, or not-for-profit sectors.

**Availability of Data and Materials**

No data was gathered or produced in current study.

**Conflicts of Interest**

None of the authors has any conflict of interest to disclose.

**Acknowledgements:** Varda and Boaz Dotan for intelligent discussion; Michal Shor for graphic design, Deborah O'hana.

**Abstract.** Quantum mechanics (QM) is derived based on a *universe composed solely of events*, for example, outcomes of observables. Such an event universe is represented by a dendrogram (a finite tree) and in the limit of infinitely many events by the p-adic tree. The trees are endowed with an ultrametric expressing *hierarchical relationships between events*. All events are coupled through the tree structure. Such a holistic picture of event-processes was formalized within the *Dendrographic Hologram Theory* (DHT). The present paper is devoted to the *emergence of QM* from DHT. We used the generalization of the QM-emergence scheme developed by Smolin. Following this scheme, we did not quantize events but rather the differences between them and through analytic derivation arrived at *Bohmian mechanics*. Previously, we were able to embed the basic elements of general relativity (GR) into DHT, and now after Smolin-like quantization of DHT, we can take a step toward quantization of GR. Finally, we remark that DHT is *nonlocal in the treelike geometry*, but this nonlocality refers to relational nonlocality in the space of events and not Einstein's spatial nonlocality.

## I. INTRODUCTION

One of the main problems in the unification of quantum and classical mechanics and general relativity (GR) is the *quantization scheme* that is applied to classical observables (see section II for the extended discussion).

The important lesson of quantum foundational studies is that quantum mechanics (QM) is about events, namely, the outcomes of measurements (*phenomena* in Bohr's terminology, [1,2]). Therefore, QM can be treated as a special formalism for event representation of physical processes (see, for example, several studies [3–10]). This is the basic idea of *the relational* *interpretation of QM* (due to Rovelli). Thus, one can try merging quantum theory and general relativity as two event-based theories [5–7,9].

From our viewpoint, one of the obstacles for event reconstruction of QM is the common use of spatial representation and Cartesian coordinates. Even Smolin who tried to consistently develop the event QM started with the spatial coordinates [6].

In recently developed *Dendrographic Hologram Theory* (DHT) as described in several studies [11–13], we proceeded with the reconstruction without using the spatial or even the temporal picture, at least as the theory's starting point (later the spatial and teporal coordinate description can emerge from the purely event-based theory [13]). The five cornerstones DHT relies on are as follows:

1. Leibniz principle (identity of indiscernibles) which states that If, for every property F, object x has F if and only if object y has F, then x is identical to y.
2. Relational and event interpretations of physical theories
3. Bohr's viewpoint on outcomes of measurements , where according to Bohr the outcomes of measurements are not the objective properties of systems. They quantitively represent interrelation between a system S and an observer O (measurement device of O).
4. Ontic-epistemic structuring of scientific theories. **Ontic description is observer independent description of reality - as it is.** Epistemic description is based on knowledge which O can extract within experiments. The ontic theory is not verifiable experimentally. It is unapproachable by the observer; the observer constructs its approximate epistemic representation of the ontic description by collecting data.
5. P-adic (ultrametric) theoretical physics

In DHT, events (Bohr's phenomena) are portrayed as branches of a dendrogram (finite tree). For an illustration that describes how to construct dendrogramic tree from data please see figure 1. These finite trees, which are the observer epistemic description of realtity, are constructed as follows: First we collect data by preforming measursments and we apply a heirarchical clustering algorithm. Then by choosing a distance matric and a certain clustering (linkage function) algorithm we construct a Agglomerative hierarchical cluster tree. In this heirarchical tree each event, which have a unique branch, can be represented by a binary string or p-adic expansion of yes/no questions. The set of branches of the tree or the strings that fully describe them is the dendrogram. Each branch runs from root to leafs, we call edges, that correspond to each event we measured. The longer the common path from root to leaf (edge) in two such branches we have a closer relation between these two events under the pre-chosen distance metric and clustering algorithm .

Within the limits of an infinite number of events, the ontic description of the event-universe is portrayed as an infinite tree. The simplest class of such trees are p-adic trees, which are homogeneous trees with p >1 edges branched from any vertex. Such trees can be endowed with the algebraic structure and the topology consistent with this structure; the p-adic topology is given by the p-adic ultrametric, namely, the strong triangle inequality holds [14]. Thus, the Dendrographic Hologram Theory (DHT) is part of well-established branch of theoretical physics, p-adic (non-Archimedean) physics [14–22], which was widely explored in string theory, cosmology, GR, theory of complex disordered systems (including spin glasses which were studied with p-adic methods by Parisi et al. [23] and others).

P-adic distance between two branches of the tree is determined by their common root in which a longer common root represents a shorter distance. *Branches represent events*, so the space of events, a finite or infinite tree, is endowed with a p-adic ultrametric. Hence, DHT does not deal with space–time localization of events but with p-adic distance encoding hierarchic relations between events. This common root distance determines *the degree of similarity between events.*

The p-adic field is endowed strange geometry where, for instance, all triangles are isosceles. This is a consequence of the strong triangle inequality. Moreover upon defininig "open" and "closed" balls as  $B(R, a) = \{ x: r_p(a, x) <= R\}$ and $B-(R, a) = \{ x: r_p(a, x) <R\}$ both balls are at the same time closed and open sets of the metric space, as such each point in a ball can be selected as its center. Geometrically a ball is a batch of infinite branches having the finite common root.

Such spaces are disordered, totally disconnected, and having zero topological dimension. However, such geometries arise very naturally from data series with application of clustering algorithms. For an illustration that describes how to construct dendrogramic tree from data please see figure 1.

Based on extensive numerical simulations DHT has been successful in simplifiying nontrivial concepts. For instance the identification of the bohmian explicate-implicate order with the dendrogramic epistemic-ontic description of universe. Moreover this connection of the explicate-implicate order with its epistemic-ontic showed that each explicate level contains the information of the lower explicate order and vice versa, Thus suggesting a holographic principle contained in each explicate order on the implicate order. The illusive bohemian quantum potential reduces in DHT to a simple structural property of the dendrogramic relational structure. Dendrogram representation of classical data violated CHSH inequality in extensive simulation study [12]. Moreover, it was shown that the hiden hierarchicl relations are non-ergodic. These simulations demonstrated that The degree of "quantumness" is checked by the violation of Bell type inequalities. The degree of classicality is based on a system's complexity, meaning the size and topological complexity of its dendrogram representation. Quantum systems are characterized by low complexity of their dendrogram topology. Another study again with extensive numerical simulations, represented GR dynamics along geodesics in a relational p-adic way [13]. Dynamics on dendrograms were emergent from simple action principle. This study also showed emergent p-adic coordinates, from the action principle, that correspond well with real spacetime coordinates.

DHT is not classical or quantum in the conventional sense. Connection with classical theory (in the general relativity (GR) framework) was established in article [13] and will be discussed in more detail in section III. At this point, we want to establish connection between DHT and QM and to emerge QM from DHT (see references [5,6,24–29] for various suggestions for emergent QM).

Based on a scheme developed by Smolin [5], we demonstrate that we can emerge QM from the event model given by DHT. Using this scheme for each event i, a set of views $V_i^k$ from i to other events k=1,…N, k≠i, is defined in which N is the total number of events. In this situation, the views rather than the spatial coordinates are the basic quantities. The view $V_i^k$ encodes the difference between the events, i and k. This kind of quantization is not quantization of positions and momenta but of differences between events. In particular, event-momentum is expressed via views. However, in Smolin's approach, the Cartesian coordinates still play a crucial role. In DHT, we start directly with the event representation given by a dendrogram. The views are determined by the hierarchic relational structure of the event-tree and differences between events are also with respect to this structure. So, our approach is more consistent as it solely explores the event picture without a direct connection to the space-time aspect. On the other hand, although we used different mathematical apparatus, we perfectly matched Smolin's ideology of event-physics.

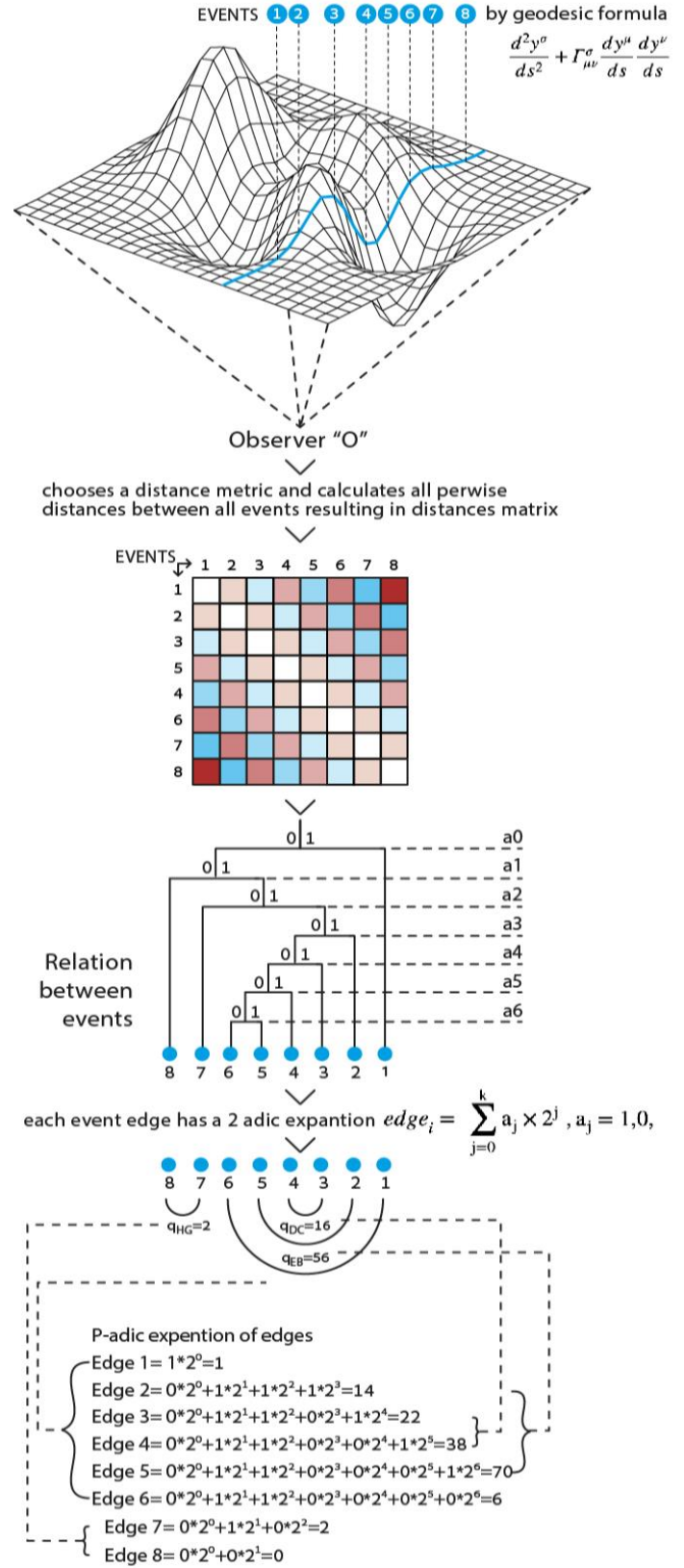

Fig 1. Illustration that describes the construction of a dendrogramic tree and its 2-adic representation from a single discrete geodesic with 8 events. A. An observer "O" measures discreate dynamics along a spacetime geodesic .B choosing a distance metric he calculates pairwise distances between the 8 discrete events along a geodesic and constructs a distance metric .
C by applying a hierarchical clustering algorithm on the distance matrix he constructs an Agglomerative hierarchical cluster tree. D in this heirarchical tree each event, which have a unique branch, can be represented by a binary string or p-adic expansion. E A numerical example is given for calculating the sum of 2-adic expansions of each event 2-adic representation.
More over 2-adic differences between two events, which are marked as $q_{ik}$, are shown as a numerical example.

We will generalize his quantization scheme to match DHT. This process allows us to use any form of distribution on the space of views and many differences/relational functions acting on the differences of events. In article [5], a very special distribution was in the use and its properties played crucial role in generating the quantum potential in the Bohmian form.

Emergence of QM in the Bohmian mechanics form immediately rises the issue of *nonlocality* and its meaning in our theory. DHT is fundamentally non-local since all events are coupled via the hierarchical relational tree-like structure. This coupling is especially evident in DHT-dynamics [30] in which appearance of a new event generates reconstruction of the whole dendrogramic universe via a recombination of the tree branches. However, this non-locality is not the same as Einstein's space-time non-locality rather it is *relational nonlocality*. As is shown in article [13], starting with the dendrogram, one can reconstruct (but not uniquely) space-time representation. In that study, the relational non-locality would be expressed in the form of apparent spatial non-locality.

## II. CANONICAL QUANTIZATION SCHEME

This section is aimed at reminding the reader that although QM functions well in numerous applications, it has some foundational problems. The reader who is not surprised by this statement can omit this section.

The quantum mechanical scheme was and still is very productive, and its consistent application led to successful development of quantum theory. One of its main advantages is the coupling to classical phase space mechanics and theory of Hamiltonian equations. This coupling in the form of deformation quantization leads to the *correspondence principle*, which is so important for heuristic justification of the quantization scheme. Deformation quantization is a rigorous mathematical procedure.

The quantum algebra of pseudo-differential operators and functions of operators of position and momentum within the limits of h to 0 coincide with the classical symbol algebra of these operators, namely, the algebra of functions on the phase space.

At the same time, it is well known that the quantization procedure has some problems that seem to be unsolvable, at least within the internal framework of QM. One such problem is the impossibility of constructing a relativistic theory for QM for which a relativistic theory was built only for quantum field theory (QFT). However, QFT suffers from infinities. Although this problem can be smoothed via numerous renormalization procedures, and these procedures are not so natural based on the heuristic viewpoint. Nowadays the problem of quantum field theory (QFT) infinities is practically ignored, but Dirac was so disturbed by it during his lifetime that he permanently demanded that the theory based on the procedure of canonical quantization is inconsistent and it should be replaced with a totally new theory [31] (see [32] for a discussion of this issue).

We note that the basic notions of quantum information theory, such as entanglement, are naturally formulated within the QM framework but not within the QFT framework [32].

Now, we point to difficulties in quantization of GR. It seems that GR is not quantizable, at least in a straightforward manner. In fact, the failure in unification of QM (or QFT) with GR is really the big black cloud on the sunny quantum sky.

We also recall the almost forgotten paper of Zeilinger in which he stressed that QM was not derived from some physically natural fundamental principles, such as the principles of special relativity, namely, the constancy of the light velocity or the relativity principle [33]. Zeilinger considered the absence of the fundamental principles as one of the main difficulties in foundational justification of QM and QFT [33].

Finally, we turn to deformation quantization and point out that this excellent mathematical theory has a big interpretational problem, namely, the treatment of the Planck constant h as a variable parameter.

The aforementioned problems led to a variety of attempts to emerge quantum theory from heuristically more acceptable theories [24–29].

### III. QUANTIZATION OF EVENTS OR VIEWS

We claim that in the event-universe, in which each event is unique, the process of quantization of concrete events (and not differences between them, namely, the views) would lead to inconsistency in the GR. The reason is that from a unique-events universe, the only available non-coarse-grained distribution is the trivial distribution. A N event universe will generate a probability distribution for which each event probability to occur is 1/N. In article [5], this distribution appeared in equation 22, but totally different and very special distribution then appeared in the formula 24 which imposes constrains on the possible configuration space. The latter leads to the right expression for the quantum potential, while QM cannot emerge with uniform distribution

It is evident that in a unique-events theory such as GR, the Leibnitz principle [11], *the identity of indiscernible*, is maintained. Practically, *this principle argues that if there are two or more identical events, they are in fact the same event.* This concept in our understanding is the main problem in uniting QM and GR. QM allows many events that are the same to construct a probabilistic theory which will have other distributions then the trivial,1/N, one as mentioned above. Therefore, we should allow many identical events to concurrently have the same view (or "they agree on the view") of the rest of the universe; in return, this process contradicts the relativity element in GR. A way out of this situation exists in which GR allows identical, or ratherm non-unique, differences between events. Thus, in GR we can construct the distributions of differences. In return, this step might lead us to an emergent QM compatible with GR.

In article [13], GR was embedded into DHT. In this article, GR's geometry was represented by its geodesics. Temporal discretization of geodesics generated the data set, which was transformed into a dendrogram. Its branches encode events, which are the points of discretized geodesics. The DHT-dendrogram expressed all relationships between events, including the space–time structure corresponding to the concrete GR-metric. By quantizing DHT resulting in the emergence of QM (via the scheme described in article [5]), we actually quantize GR event data via its DHT representation. Of course, these schemes are just schematic manipulations, which are far from the rigid basis of quantum gravity.

## IV. EMERGENCE OF QUANTUM REPRESENTATION FROM DH-THEORY

### A. The fundamental "difference pdf"

We follow the line developed by Smolin [5] with slight modifications and conversion to our DH-theory point of view.

For our purposes, we used the equation shown below:

$edge_i = \sum_{j=0}^{k} a_j \times 2^j$ is a dendrogram branch 2-adic expansion. Its monna map conversion is defined as: $event_i = \sum_{j=0}^{k} a_j \times 2^{-j-1}$, $a_j = 1,0,$ (1)

$q_{ik} = (event_i - event_k)$ thus $V_i^k = q_{ik}$ in which $k$ is any of the remaining $N-1$ edges and $k \neq i$

We then define the measure of differences called distinctiveness, same as in [5]:

$$I_{ij} = \frac{1}{N}\Sigma_k(V_i^k - V_j^k)^2 \qquad (2)$$

Thus, the larger $I_{ij}$, the more easily event$_i$ and event$_j$ can be differentiated by their views. We then defined the variety as shown below, same as in [5]:

$$v = \frac{A}{N^2}\Sigma_{i \neq j} I_{ij} \qquad (3)$$

Or as inter ensemble potential energy.

We start with constructions of the "differences pdf", $\rho$. This will be our fundamental distribution. Thus, given a dendrogram that describes, p-adicaly, the relations between m events we calculate all possible pairwise differences of the p-adic edges representation and represent them as events through monna map as above:

$$q_{ik} = (event_i - event_k) \qquad (4)$$

Where for p = 2,

$$edge_i = \Sigma_{j=0}^{k} a_j \times 2^j \rightarrow event_i = \Sigma_{j=0}^{k} a_j \times 2^{-j-1}, a_j = 1,0, \qquad \text{as in (1)}$$

the Monna map maps natural numbers into rational numbers belonging the segment [0,1]. A dendrogram is mapped into a subset of [0,1]. This map can be extended to the infinite p-adic tree where its branches are represented by infinite series; for p=2,

$$edge_i = \Sigma_{j=0}^{\infty} a_j \times 2^j \rightarrow event_i = \Sigma_{j=0}^{\infty} a_j \times 2^{-j-1}, a_j = 1,0, \qquad \text{as in (1)}$$

The latter is important for considering the limit of the universe with infinitely many events; they are portrayed on an infinite 2-adic tree.

Thus, from all $q_{ik}$ we have a discrete "difference pdf", $\rho$, which is the fraction of each unique value of $q_{ik}$ defined For the set of the unique $q_{ik}$ values Q we have:

$$\rho_j = \frac{(number\ of\ different\ q_{ik}\ that\ equal\ Q_j)}{total\ number\ of\ q_{ik}}$$

$$(5)$$

### B. The differences energy

To define the differences energy, which will be equivalent to the usual kinetic energy, we define the equation:

$$p = \frac{1}{N}\sum_{k \neq j} q_{jk} \text{ in which N is the number of possible } q_{jk}. \tag{6}$$

This formula represents the mean difference between all pairwise events. Thus, $p$ takes the part as the momentum in Bohmian mechanics.

Now, we introduce the phase $S$ as the new parameter of the model:

We set $p = \partial S$ and $\partial S = e^{iS}$. $\tag{7}$

We can then calculate the "differences energy" as the real part of:

$$T_{differences\ energy} = \frac{1}{N}\sum_{j \neq k}(q_{jk})^2 \tag{8}$$

Where $T_{differences\ energy}$ serves as our equivalent of kinetical energy

The continuum approximation of $T_{differences\ energy}$ will be obtained by noticing

$$p_j = \rho(Q_j)(Q_j)^2 = \partial S_j \tag{9}$$

Thus:

$$T_{differences\ energy} = \int (\rho(Q_j)(Q_j)^2)\rho(Q_j)dQ_j = \int (\partial S_j)^2 \rho(Q_j)dQ_j \tag{10}$$

### C. The continuous dendrogram differences energy

The dendrogramic dynamical change from D1 to D2…Dn

Requires another term in the action, we call, **dendrogram differences energy.**

$$T_{dendrogram\ differences\ energy} = \int d(Dendrogram) \sum_j \frac{dp}{dDendrogram} \tag{11}$$

And since

$$p = \rho(Q)Q = \partial S$$

$$T_{dendrogram\ differences\ energy} = \int d(Dendrogram) \int \frac{d\int \rho(Q)(Q)^2 dQ}{dDendrogram}\rho(Q)dQ =$$

$$= \int d(Dendrogram) \int \frac{d(S)}{dDendrogram}\rho(Q)dQ \tag{12}$$

Thus, our fundamental action would be the same as Smolins

$$A(S, Q) = \int d(Dendrogram)\{\int \frac{d(S)}{dDendrogram} \rho(Q)dQ + \int (\partial S)^2 \rho(Q)dQ - v(Q) + U(Q)\}$$

(13)

### D. The origin of the quantum potential

We started with $V_k^i = (edge_i - edge_k)$ (14)

we identify all $V_k^i$ that have the same value as Q

with probability $\rho(Q)$ to occur in the dendrogram

Thus, the expected $V_k^i$ is for all i's and k's

$$E[V_k^i] = \int Q\rho(Q)dQ$$

(15)

now, the veriaty is as above $v = \frac{A}{N^2}\sum_{i \neq j} I_{ij}$ where $I_{ij} = \frac{1}{N}\sum_k (V_i^k - V_j^k)^2$

Or in the continuous form

$$v = \int dQ\rho(Q)Z_V \int dx \int dy [((Q+x) - (Q+y))^2 \rho(Q+x)\rho(Q+y)]$$

(16)

Thus, we have

$$\int dQ\rho(Q)Z_V \int_0^1 dx \int_0^1 dy[(x-y)^2 \rho(Q+x)\rho(Q+y)]$$

(17)

which is equivalent to $Z_V \sum_{j \neq k} \sum_k \sum_{i \neq j} (q_{jk} - q_{ik})^2$ where $Z_V$ is a normalization factor.

We show now the emergence of the quantum potential term in our action where $\rho(Q+x)$ and $\rho(Q+y)$ are expended in x and y around Q. we take the two integral terms with $(\rho(Q))^2$ and $(\partial \rho(Q))^2$. Thus, we obtain two equations:

$$Z_V(\rho(Q))^2 \int_0^1 dx \int_0^1 dy(x-y)^2 = \frac{1}{6} Z_V(\rho(Q))^2$$

$$Z_V(\partial\rho(z))^2 \int_0^1 dx \int_0^1 dy(x-y)^2 xy = \frac{1}{36} Z_V(\partial\rho(Q))^2$$

(18)

when inserted back into (17):

$$\int dQ \rho(Q) Z_V \left(\frac{1}{6}(\rho(Q))^2 + \frac{1}{36}(\partial\rho(Q))^2\right) + \frac{1}{120}(\nabla^2\rho(z))^2 \ldots$$

with $Z_V = -\frac{36}{(\rho(Q))^2}$

(19)

$$\int dQ \rho(Q)\left(-6 - \frac{(\partial\rho(z))^2}{(\rho(Q))^2}\right) - 36/120 \frac{((\nabla^2\rho(z))^2)}{(\rho(Q))^2} \ldots \ldots)$$

(20)

Here we ignore total derivatives, which don't contribute to the potential energy. The first term is an ignorable constant. The second term is what we want; its variation gives the Bohmian quantum potential and The leading correction is which contributes non-linear corrections to the Schroedinger equation.

The full continuous action is thus:

$$A(S,\rho) = \int d(Dendrogram)\{\int \frac{d(S)}{dDendrogram} \rho(Q) dQ + \int (\partial S)^2 \rho(Q) dQ - v(Q) + U(Q)\}$$

Where $v(Q) = Z_V \int \rho(Q)\left(\frac{(\partial\rho(z))^2}{(\rho(Q))^2}\right) dQ$ and $U(Q) = \int \rho(Q) dQ$

(21)

Where $U(Q) = \int \rho(Q) dQ$

The Hamilton Jacobi equation of the action (notice we are on the straight line [ 0 1] now in all our pdfs ).

$$-\dot{S} = (\partial_{edge} S)^2 + U + U^Q \qquad (22)$$

In which $U^Q = (\Delta^2 \sqrt{\rho})/\sqrt{\rho}$ is the quantum potential

And $U$ is the potential

And $\dot{S} = S(present\ dendrogram) - S(previous\ dendrogram)$

All parameters are well-defined.

The probability conservation law is

$$\dot{\rho} = \partial_{edge}(\rho\partial_{edge}S)^2$$

In which $\dot{\rho} = \rho(present\ dendrogram) - \rho(previous\ dendrogram)$ (23)

Equations 22 and 23 are the real and imaginary parts of the Schrodinger equation for $\psi = \sqrt{\rho}e^{iS}$. please notice the arguments in the action must follow least action principle thus they are all following best match (or best distance) distance between two zero dimensional dendrogram structures which is with agreement with particle shape dynamics theory. In our case the particles are replaced with events encoded by dendrograms p-adically. Thus, $\rho$ is the probabilistic representation of the dendrogramic structure where events correspond to particles in shape dynamics. $S$ is again a property of the events encoded in the dendrogram. Thus, we closely linked casual set theory with particle shape dynamics under our formulation with an emergent QM. Notice that shape dynamics do not yet have a quantum theory.

We now turn to the link with the gravitational field as stated above.

We use as our probabilistic representation (or our quantum representation), the Hamiltonian:

$$H = \rho(z)[(\partial S)^2/2 - (\frac{\partial \rho(z)}{\rho(z)})^2 + U]$$

With the canonical variables $\rho$ and $S$ and where $U = \rho$. Please note that $\rho$ represents the variable $edge$, and $S$ the variable $\int p$ where $p$ is defined as $\frac{1}{N}\sum_{k \neq j} edge_j - edge_k$.

but in our construction of the dendrogram we have used a clustering function so

$\rho(edge) = F(\frac{d^2x^\sigma}{ds^2} + \Gamma^\sigma_{\mu\nu}\frac{dx^\mu}{ds}\frac{dx^\nu}{ds})$ where $F$ is a function which clusters spacetime coordinates, x, by some algorithm and then gives each single cluster a p-adic representation. x are spacetime coordinates . The procedure applies also to

$S(edge) = \int\int\int F\left(\frac{d^2x^\sigma}{ds^2} + \Gamma^\sigma_{\mu\nu}\frac{dx^\mu}{ds}\frac{dx^\nu}{ds}\right) - F\left(\frac{d^2y^\sigma}{ds^2} + \Gamma^\sigma_{\mu\nu}\frac{dy^\mu}{ds}\frac{dy^\nu}{ds}\right) dxdyd(edge)$ and where

y=x+z and x and y are spac-etime coordinates

In which $\Gamma^\sigma_{\mu\nu}$ is the gravitational force.

## V. DISCUSSION

DHT [11–13,30] presents the event picture of the universe. This picture does not contain the physical space, rather, it has the treelike geometry expressing hierarchical relationships between events. The relational structure generates a type of nonlocality effect. This effect is especially evident in the dynamical model for DHT [30] in which the appearance of a new event induces recombination of all events on the tree and interrelation between them.

The DHT model is not classical or quantum in the sense of conventional physics. The discrete structure of branching on the event-tree suggests quantization. In article [12], we found some quantum-like effects within DHT, for example, the possibility to violate the Bell type inequalities, which were treated as the tests of the level of quantum properties.

However, coupling of DHT with standard QM was missing up to that point. This coupling was established in the present paper based on Smolin's scheme for emergence of QM from the view-universe. Although ideologically Smolin advertised the purely event pictured universe, in concrete calculations he needed the standard mathematical apparatus with real Cartesian coordinates, at least in article [5]. We start without space–time, and just used a hierarchical tree of events. However, we found that the Smolin quantization scheme can be generalized for state space of a dendrogram within the limitations of the p-adic type.

One of the aims of quantization of DHT is to take a step toward quantization of GR, which was coupled to DHT in article [13]. As explained in sections II and III, the individual structure of events in GR cannot be expressed in the probabilistic framework of QM (see figure 2) . Therefore, to emerge QM from DHT, we should try not to operate with events distributions. Following Smolin, we operated with distributions of differences between events expressed in terms of views from one event to another.

We can say that the mission of DHT quantization was successfully completed. Of course, this completion is just the first step. In future work, we should represent the basic DHT quantities in standard quantum terms. One of interesting problem is construction of the quantum representation of correlations between views.

Further studies should address the formulation of other properties of QM in our the DHT approach such as the emergence of spin and the characterization of different field types along with their corresponding QM properties. Moreover, we speculate that the connection to shape dynamics, in terms of dynamics of relations structures would increase our predictive powers of the dynamical processes in the event universe

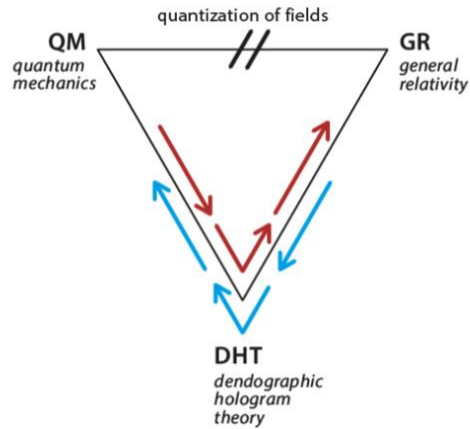

Fig 2. An abstract illustration that suggests DHT can describe quantum phenomena, general relativity phenomena and quantize gravitation field. Direct quantization from GR to QM is still missing.